\documentclass[preprint,aps,12pt,preprintnumbers,eqsecnum,nofootinbib]{revtex4}
\usepackage{graphicx}
\usepackage{subfigure}

\usepackage{color}
\usepackage{amssymb,amsmath}

\newcommand{\beq}{\begin{equation}}
\newcommand{\enq}{\end{equation}}

\begin{document}
%
%
\title{\vspace*{0.5in} 
Deconstructing Superconductivity
\vskip 0.1in}
\author{Dylan Albrecht}\email[]{djalbrecht@email.wm.edu}
\author{Christopher D. Carone}\email[]{cdcaro@wm.edu}
\author{Joshua Erlich}\email[]{jxerli@wm.edu}

\affiliation{High Energy Theory Group, Department of Physics,
College of William and Mary, Williamsburg, VA 23187-8795}
\date{August 2012}
\begin{abstract}
We present a dimensionally deconstructed model of an $s$-wave
holographic superconductor.  The 2+1 dimensional model includes
multiple charged Cooper pair fields and neutral exciton fields that 
have interactions governed by hidden local symmetries. We derive 
AdS/CFT-like relations for the current and charge density in the model, 
and we analyze properties of the Cooper pair condensates and the 
complex conductivity.
\end{abstract}
\maketitle

\section{Introduction} \label{sec:intro}
High-$T_c$ superconductors continue to fascinate physicists,
both for their technological prospects and for their unconventional physical properties.
Discovered in 1986~\cite{Bednorz:1986tc}, high-$T_c$ 
superconductors have superconducting transition temperatures $T_c$ higher
than the upper limit of around 30~K suggested by the BCS theory~\cite{McMillan}, and the
mechanism of superconductivity in these materials remains uncertain.
High-$T_c$ superconductors are Type II, allowing magnetic fields to penetrate
in conjunction with Abrikosov vortex currents which maintain localized magnetic fluxes.
A number of features of high-temperature superconductors
remain poorly understood, including the existence of a pseudogap phase in which a 
gap in the excitation spectrum opens up at temperatures above the superconducting 
transition~\cite{pseudogap}, and a thermoelectric Nernst effect that occurs in both the
superconducting and pseudogap phases~\cite{Nernst}. A few additional classes of exotic 
high-$T_c$ superconductors have recently been discovered, including the iron pnictides~\cite{pnictides},
and research into their properties is ongoing.

In the absence of a clear theoretical understanding of high-$T_c$
superconductors and other systems with strongly correlated electrons,
it is useful to explore simplified models which describe similar
phenomenology.  It has long been understood that certain relativistic field
theories describe a variety of phenomena typical of superconductors.  In
the Abelian Higgs model, a charged scalar field condenses while the photon
becomes massive, in analogy to the condensation of Cooper pairs
and the consequent expulsion of magnetic fields from superconducting
materials.  The Abelian Higgs model also has Nielsen-Olesen
vortex solutions which maintain localized magnetic flux tubes~\cite{Nielsen-Olesen}, much like
the Abrikosov vortices in Type II superconductors.

One approach that has stimulated much effort especially
from the string theory community is the use of the AdS/CFT
correspondence~\cite{AdSCFT}, which relates certain strongly coupled 
field theories to gravitational systems in higher-dimensional spacetime backgrounds.
Due to the extra-dimensional nature of these
models they are referred to as holographic.  There is a mapping between
observables in the superconducting system and
fields in the corresponding extra-dimensional model.
Progress has been made in understanding properties of holographic models of
superconductors and other condensed matter systems,
including $s$, $p$, and $d$-wave superconductivity~\cite{Gubser,HHH,holo-pd-wave}, the
Nernst effect in the quantum critical regime~\cite{holo-Nernst}, strange metallic 
behavior~\cite{strange-met}, enhancement of the 
superconducting gap compared to BCS theory~\cite{Horowitz-rev},
and Homes's empirical law for $T_c$ as a function of low-temperature superfluid density
and normal-phase resistivity~\cite{Homes,holo-Homes}.  In some models of holographic superconductors
there is no hard zero-temperature gap~\cite{Horowitz-Roberts}.  For some reviews
on holographic models of condensed matter systems, see Refs.~\cite{Horowitz-rev,Herzog-rev,Hartnoll-rev}.

In parallel to the development of the AdS/CFT correspondence in string
theory, extra-dimensional model building has led to
the discovery of new paradigms for gravitational physics~\cite{ADD,RS}, 
electroweak symmetry breaking~\cite{xdEWSB}, grand
unification~\cite{xdGUT}, and other possible new physics.
Notably, in order to define these otherwise nonrenormalizable
extra-dimensional gauge theories, a procedure was developed
for dimensionally ``deconstructing'' extra-dimensional theories into renormalizable
field theories defined in the natural (lower) dimension of the system of
interest~\cite{decons}. At long distances, a deconstructed model is effectively described by
the corresponding extra-dimensional theory, latticized in the extra dimension(s).
In some models the number of effective lattice sites can be
taken as small as just two or three while retaining the desirable
features of the extra-dimensional theory.  Deconstruction provides a systematic 
procedure for finding relatively simple, weakly coupled models of interesting physical 
systems, in the number of dimensions that naturally characterizes the system 
of interest~\cite{littleHiggs}. 

In this paper we present and analyze a deconstructed model of the simplest holographic
superconductor, the Abelian Higgs model in 3+1 dimensional\footnote{The +1
in the spacetime dimensionality always refers to time, not the extra spatial
dimension of the holographic model.}
Anti-de Sitter-Schwarzschild spacetime (AdS$_4$-Schwarzschild), which is intended to model
$s$-wave superconductors in 2+1 dimensions~\cite{HHH}.
Superconductivity in the cuprates is along two-dimensional CuO$_2$ planes, so the
hope is that a model with two spatial dimensions 
will capture some of the relevant physics in these systems.  
The deconstructed model is a 2+1 dimensional model in which Maxwell's 
electrodynamics is replicated a number of times.
Fields charged under pairs of the electromagnetic U(1) gauge groups condense, 
giving rise to the latticized extra-dimensional structure of the model.
The temperature-dependent couplings of the Maxwell fields are
rotationally invariant but not Lorentz invariant.  Additional charged
fields representing Cooper pair operators condense in the
superconducting phase, and the combined effects of
the multiple condensates determine the properties of the superconductor.
The electron pair wavefunctions
in high-$T_c$ superconductors are thought to have more complicated symmetry than the $s$-wave 
described by scalar Cooper pair fields, so the present model is not expected to capture 
features sensitive to the symmetry of the wavefunctions.

The deconstructed model helps to elucidate certain aspects of holographic 
superconductivity and suggests the relevant effective degrees of freedom
in physical realizations of these superconductors.  One generic feature of
this class of models is the existence of hidden local symmetries~\cite{HLS} and corresponding
neutral excitations, much like the vector mesons of Quantum Chromodynamics.  Such
excitations in the superconductor may be interpreted as excitons\footnote{The neutral, spin-1 
excitations might more appropriately be identified with polaritons, if there were a dynamical photon in
the deconstructed theory that mixed strongly with these excitons.}, 
electron-hole bound states which can be created when a photon
is absorbed by certain materials~\cite{exciton}. In analogy to a deconstructed holographic model of
hadrons~\cite{Son-Stephanov}, we derive discretized AdS/CFT relations between bulk 
and boundary observables. We analyze properties of the superconducting transition and the
frequency-dependent conductivity as the number of extra-dimensional lattice sites, and 
correspondingly the number of pair condensates and exciton fields, is reduced.

In Sec.~\ref{sec:cmodel}, we review the Abelian Higgs model
in the AdS$_4$-Schwarzschild spacetime as a model of superconductivity in
two spatial dimensions.  In Sec.~\ref{sec:moose}, we describe the
deconstructed model and explain the calculation of  observables of interest.  In
Sec.~\ref{sec:results}, we present numerical results.  We conclude with a discussion
and suggestions for future research in Sec.~\ref{sec:conclusions}.

\section{Continuum Model} \label{sec:cmodel}
In this section we review the holographic model that will be deconstructed in 
Sec.~\ref{sec:moose}.  Much of the content of this section is a summary of
results contained elsewhere, for example in Ref.~\cite{HHH}, though we generalize 
certain results for the sake of comparison with the deconstructed model.

The starting point is a vacuum solution to Einstein's equations with
negative cosmological constant in 3+1 dimensions, namely the
AdS$_4$-Schwarzschild solution.  The spacetime is described by a metric
of the form \beq
ds^2=F(r)dt^2-r^2(dx^2+dy^2)-\frac{1}{F(r)}dr^2, \label{eq:AdS-Sch-r}\enq
where \beq
F(r)=\frac{r^2}{L^2}\left(1-\frac{r_H^3}{r^3}\right). \label{eq:F-r} \enq
The constant $L$ is the Anti-de Sitter scale, which we will often set to 1.
We refer to coordinates in which the metric takes
the form of Eq.~(\ref{eq:AdS-Sch-r}) as $r$ coordinates, and the coordinate
$r$ runs from $r_H$ to $\infty$.  The 2+1 dimensional surface at
$r=\infty$ is referred to as the ultraviolet (UV) 
boundary, and the surface at $r=r_H$ is the
horizon.  The AdS/CFT correspondence suggests the identification of the temperature
of the material with the Hawking temperature of the AdS black hole, namely
\beq T=3r_H/(4\pi L^2). \enq
With nonvanishing charge density the relevant background spacetime is
instead the AdS-Reissner-Nordstrom spacetime, a charged solution to the 
Einstein-Maxwell equations with a negative cosmological constant. However, 
in this paper we neglect the backreaction of the charge density
on the geometry.  In this approximation the spacetime metric continues
to take the form of Eqs.~(\ref{eq:AdS-Sch-r}) and (\ref{eq:F-r}), even when
the solutions of interest have nonvanishing charge density.  

The holographic model contains a U(1) gauge field corresponding to
electromagnetism, and a charged scalar field corresponding to the Cooper pairs.  The 
model is translationally invariant in 2 spatial dimensions, and is therefore not expected to 
reproduce phenomena that are sensitive to the atomic lattice.

The action for the model is given by \beq
S=\int d^4x\,\sqrt{g}\left\{-\frac{1}{4}F_{MN}F^{MN}+|(\partial_M-i A_M)\psi|^2
-m^2|\psi|^2\right\},
\enq
where index contractions are with the 4D metric $g_{MN}$, and $g=|{\rm det}\,g_{MN}|$.

The action in $r$ coordinates has the form \beq \begin{split}
S=&\int d^3x\,dr\,r^2\left\{\frac{1}{2r^2 F(r)}(F_{0a})^2-\frac{1}{4}
r^{-4}(F_{ab})^2
+\frac{1}{2}(F_{0r})^2-\frac{F}{2 r^2}(F_{ar})^2 \right.\\
&\left.+\frac{1}{F(r)}|(\partial_0-i A_0)\psi|^2-r^{-2}
|(\partial_a-iA_a)\psi|^2-F(r)|(\partial_r-iA_r)\psi|^2-m^2|\psi|^2\right\}.
\end{split}
\label{eq:r-action}\enq
Here $F_{MN}=\partial_MA_N-\partial_NA_M$ is the field strength of the
U(1) gauge field, and the charge of the field $\psi$ is normalized to 1.
Our convention is that Latin indices $a,\ b,$ {\em etc.}, represent spatial
components $x$ or $y$. We have explicitly shown the metric factors in Eq.~(\ref{eq:r-action}), so
remaining index contractions are taken with the Kronecker delta $\delta_{MN}$.
According to the AdS/CFT correspondence the mass $m$ is related to the
scaling dimension of the operator related to $\psi$, and for definiteness
we will assume $m^2=-2/L^2$ as in Ref.~\cite{HHH},
corresponding to a Cooper pair operator of dimension 2 (or 1, as discussed below).

Defining $\phi\equiv A_0$ as in Ref.~\cite{HHH}, and considering solutions with $\psi=\psi(r)$,
$\phi=\phi(r)$ and $A_a=0$, the coupled equations of motion for $\psi$ and $\phi$ in the gauge 
where $A_r=0$ are
\beq
\psi''+\left(\frac{F'(r)}{F(r)}+\frac{2}{r}\right)\psi'+\frac{\phi^2}{F(r)^2}\psi-
\frac{m^2}{F(r)}\psi=0,\enq
\beq
\phi''+\frac{2}{r}\phi'-\frac{2|\psi|^2}{F(r)}\phi=0.\enq
Near the UV boundary, the solutions behave as~\cite{HHH},
\beq
\psi=\frac{\psi^{(1)}}{r}+\frac{\psi^{(2)}}{r^2}+\cdots,
\enq
\beq\phi=\mu-\frac{\rho}{r}+\cdots.
\label{eq:phi-r}\enq
The time component of the gauge field couples to the charge density, so the AdS/CFT
correspondence identifies $\mu$ with the electric chemical potential and $\rho$ with the
charge density.  For the Cooper pair operator there are two choices: either $\psi^{(1)}$
acts as a source  and $\psi^{(2)}$ corresponds to the expectation value
of the Cooper pair operator, or {\em vice versa}.  
This choice determines the scaling dimension
of the Cooper pair operator according to the AdS/CFT correspondence.  
For definiteness we will choose to take 
$\psi^{(1)}$ as the source of the Cooper pair operator $O_2$, which then has mass dimension
two; the other choice would correspond to an operator of dimension 1.
The Cooper pair operator should only be turned on dynamically, so vanishing of its
source becomes a UV boundary condition, $\psi^{(1)}=0$.
Additional boundary conditions
are $\phi(r_H)=0$ for regularity of the solution
and $\phi(\infty)=\mu$.  The equations of motion enforce the condition
$F'(r)\psi'(r)=m^2\psi(r)$ at the horizon if $\psi(r)$ is finite, so the additional boundary 
condition on $\psi$ at the horizon is replaced with a regularity condition.

In order to study the response of our system to an oscillating electric field,
we consider solutions for which $A_x$ oscillates in time, $A_y=A_r=0$ and $A_0$ behaves as in 
Eq.~(\ref{eq:phi-r}).
Assuming the form \begin{equation}
A_x(t,r)=e^{-i\omega t}A(r),\end{equation}
the bulk equation of motion for $A_x$ has the form,
\begin{equation}
-\frac{\omega^2}{F(r)}A(r)-\frac{d}{dr}\left(F(r)A'(r)\right)+2A(r)|\psi(r)|^2=0.
\label{eq:AxEOM}\end{equation}
For large $r$ the solutions for $A_x$ have the form
\begin{equation}
A_x= A_x^{(0)}+A_x^{(1)}/r + \cdots .\end{equation}
According to the AdS/CFT correspondence, $A_x^{(0)}$ acts as the source for
the current $j^x$ and hence corresponds to a background
electric field $E_x=\partial_tA_x^{(0)}$.  The AdS/CFT correspondence
also suggests an identification of the normalizable
component $A_x^{(1)}$ with (minus) the current $j^x$ in the given
background state~\cite{bulk-bdry}.  In Sec.~\ref{sec:moose}, we explain how this identification 
arises in the deconstructed model. Hence, we can write\footnote{The sign differences in these
expressions compared with Ref.~\cite{HHH} are due to differences in convention related to the signature of the
metric.  We follow the conventions of Jackson's {\em Classical Electrodynamics} \cite{Jackson}.}
\begin{equation}\begin{split}
E_x&=\partial_tA_x|_{r\rightarrow\infty}=-i\omega A_x|_{r\rightarrow\infty}, \\
j^x&=-A_x^{(1)}=r^2\partial_rA_x|_{r\rightarrow\infty}. \end{split}\end{equation}
The conductivity $\sigma$ is therefore determined in the holographic model by
\begin{equation}
\sigma=\frac{j^x}{E_x}=\left.\frac{r^2 A'(r)}{-i\omega A(r)}\right|_{         
r\rightarrow\infty},\label{eq:sigma}\end{equation}

\subsection{The Normal Phase}
At temperatures greater than $T_c$ the charged field
$\psi$ vanishes, in which case
solutions of the following first order equations also solve
Eq.~(\ref{eq:AxEOM}):  \begin{equation}
A_{\pm}'(r)=\pm \frac{i\omega}{F(r)}A_{\pm}(r).\label{eq:Afirstord}\end{equation}
The solutions are \begin{equation}\begin{split}
A_{\pm}(r)&=\exp\left\{\pm i\omega\int \frac{dr}{F(r)}\right\}\\
&=\exp\left\{\pm \frac{i\omega}{6 r_H}\left(2\sqrt{3}\tan^{-1}\left[\frac{2r+r_H}{
\sqrt{3}r_H}\right]
+\log\left[\frac{(r-r_H)^2}{r^2+r r_H+r_H^2}\right]\right)\right\},\label{eq:Aanalytic}
\end{split}\end{equation}
and the generic solution to Eq.~(\ref{eq:AxEOM})
is a linear combination $A(r)=c_+ A_+(r)+c_- A_-(r)$ with constants $c_+$ and
$c_-$.
The choice of branch of the inverse tangent and the logarithm in
Eq.~(\ref{eq:Aanalytic})
determines an $\omega$-dependent constant multiplying each of the two
solutions, which can be absorbed in the coefficients $c_+$ and $c_-$.
The solution $A_-(r)$ describes a wave flowing
into the black hole horizon, and $A_+(r)$ describes an outgoing wave.
Son and Starinets have advocated the choice of
ingoing-wave solution as a boundary condition at the horizon, based on
the requirement of causality of correlators calculated by the AdS/CFT
correspondence~\cite{Son-Starinets}.  For now, we explore the
consequences of the generic solution, for guidance as to what to expect from
the deconstructed model. A natural choice for the boundary condition in the discretized 
theory is more ambiguous because causality may be imposed at the level of the
lower-dimensional theory, without regards to the effective higher-dimensional
description.

The complex conductivity as a function of frequency distinguishes the normal 
and superconducting phases.  A hallmark of perfect conductivity is a delta function in the real
part of the conductivity at zero frequency which, by the Kramers-Kronig
relation for the conductivity, is tantamount to a zero-frequency pole in the imaginary part.  The
Kramers-Kronig relation follows from the assumption that correlations are causal.
As a consequence, the conductivity $\sigma(\omega)$, which 
by definition is the Fourier transform of the response
function relating the current to a background electric field, is
analytic in the upper half plane.  The Kramers-Kronig relation is then
the statement of Cauchy's integral theorem for the following integral:
\begin{equation}
\oint_C\frac{\sigma(\omega')}{\omega'-\omega}d\omega'={\cal P}
\int_{-\infty}^\infty\frac{\sigma(\omega')}{\omega'-\omega}d\omega'-
i\pi\,\sigma(\omega)=0, \label{eq:Kramers-Kronig}\end{equation}
where the contour $C$ spans the real axis above the pole
at $\omega'=\omega$ and closes in the upper half plane, and ${\cal P}$ represents the principal 
value of the integral.  The conductivity $\sigma(\omega)$ is assumed to fall off in the upper half 
plane faster than $1/|\omega|$, so that that integral over the contour at infinity vanishes.  As a consequence of 
Eq.~(\ref{eq:Kramers-Kronig}), if the real part of $\sigma(\omega)$ has a delta-function at $\omega=0$ then the 
imaginary part has a pole at $\omega=0$, and {\em vice versa}.

It follows from Eqs.~(\ref{eq:sigma})
and (\ref{eq:Afirstord}) that when $\psi$ vanishes, the zero-frequency pole
in Im~$\sigma$ is generically absent; the holographic description indeed
describes a non-superconducting phase when the condensate vanishes,
for a generic choice of boundary condition at the horizon.
Note that for both the ingoing-wave and outgoing-wave solutions, the
normal-phase conductivity is independent of $\omega$:
$\sigma=+1$ for the ingoing wave, and $\sigma=-1$ for the
outgoing wave.  This result provides a phenomenological  motivation for the
ingoing-wave boundary conditions: the outgoing-wave solution would
describe an unusual situation in which the
current produced by an electric field points in the direction opposite to the
electric field.  Note  also that the conductivity in this model
does not indicate the existence of a pseudogap phase, and does not 
fall off at large $\omega$ due to relaxation as in ordinary materials.

For the generic solution for $A_x$ with vanishing $\psi$ and with the identification
$r_H=4\pi T/3$, Eq.~(\ref{eq:sigma})  gives the normal-phase
conductivity: \begin{equation}
\sigma(\omega)=\frac{c_-\exp\left(\frac{-i\sqrt{3}\omega}{8 T}\right)
-c_+\exp\left(\frac{i\sqrt{3}\omega}{8 T}\right)}{
c_-\exp\left(\frac{-i\sqrt{3}\omega}{8 T}\right)+
c_+\exp\left(\frac{i\sqrt{3}\omega}{8 T}\right)}.\end{equation}
The real part of the conductivity is then, \begin{equation}
{\rm Re}\,\sigma(\omega)=\frac{|c_-|^2-|c_+|^2}{|c_-|^2+|c_+|^2
+c_-c_+^*e^{\frac{-i\sqrt{3}\omega}{4 T}}+
c_+c_-^*e^{\frac{i\sqrt{3}\omega}{4 T}}}, \label{eq:wiggles} \end{equation}
which oscillates as a function of $\omega/T$ around $\sigma=1$  for constant $|c_+| \ll |c_-|$.

\subsection{The Superconducting Phase}
For small enough $T$, the coupling between $\phi$ and $\psi$ leads to an instability which
turns on the Cooper pair condensate $\langle O_2 \rangle $.  The condensate depends on the chemical potential $\mu$ and
the charge density $\rho$ determined by the solution to the coupled equations of motion.  
As argued in Ref.~\cite{HHH}, the critical temperature scales as $\rho^{1/2}$.  Near $T_c$ the
condensate behaves as $\langle O_2 \rangle \propto(1-T/T_c)^{1/2}$.  The complex conductivity for
generic $T<T_c$ has a delta function at $\omega=0$ and a gap $\omega_g$, as expected for 
typical superconductors, but with $\omega_g/T_c\approx8$~\cite{HHH}, significantly larger than
the BCS prediction $\omega_g/T_c\approx 3.5$.  The phenomenology of the deconstructed model has similarities 
to that of the continuum theory, even with only a few lattice sites, as we will see in Sec.~\ref{sec:results}.

\section{Deconstruction} \label{sec:moose}

In the deconstructed model we will define coordinates differently,
replacing $r$ with a new coordinate $z=1/r$.  In $z$ coordinates,
the extra dimension corresponds to a finite coordinate interval.  The metric 
in $z$ coordinates has the form \beq
ds^2=\frac{1}{z^2}\left[f(z)dt^2-\frac{1}{f(z)}dz^2-(dx^2+dy^2)\right],
\label{eq:AdS-Sch-z}\end{equation}
where  \beq
f(z)=\frac{1}{L^2}\left(1-\frac{z^3}{z_H^3}\right), \label{eq:f-z}\enq
where $z$ runs from 0 to $z_H$.  The surface $z=0$ is the ultraviolet boundary,
and $z=z_H$ is the horizon.  One should keep in mind that results are coordinate
independent in the continuum model, but away from the continuum limit, deconstructed models 
depend on the choice of coordinates in which their extra-dimensional parent model is defined.

In $z$ coordinates, the action in the continuum theory takes the form
\begin{eqnarray}
S &=& \int d^4 x \left\{ \frac{1}{2} F_{0z}^2 + \frac{1}{2 f} F_{0a}^2 -\frac{f}{2} F_{za}^2 -\frac{1}{4} F_{ab}^2
+\frac{1}{z^2 f} | \partial_0 \psi -i A_0 \psi |^2 \right. \nonumber \\
&& \left.   -\frac{f}{z^2} |\partial_z \psi -i A_z \psi |^2-\frac{1}{z^2} |\partial_a \psi -i A_a \psi|^2 
- \frac{1}{z^4} m^2 |\psi|^2 \right\} \, ,
\end{eqnarray}
where repeated $a$ or $b$ indices are summed over the spatial coordinates $x$ and $y$.  As
before, we use the notation of Ref.~\cite{HHH}, where
we define $\phi \equiv A_0$.   We replace the $z$ axis by a lattice of $N$ points,
\begin{equation}
z_j = \left\{ \begin{array}{ll}
\epsilon+(j-1)a  & \mbox{for  $j=1 \ldots N-1$} \\
\epsilon+(N-2)\, a + a_H & \mbox{for $j=N$} \,\,\, , \end{array}\right. \,
\end{equation}
where $z_N = z_H$ and $\epsilon$  is a UV  cutoff.  The lattice spacing $a_j$ between the 
$j$ and ${j+1}^{\rm th}$ lattice sites is therefore $a$ for $j=1\ldots N-2$ and $a_H$ for $j=N-1$. 
We give ourselves the freedom to vary $a_H$ away from $a$ to test the sensitivity of our results
to the density of sites nearest to the horizon. The discretized action is given by
\begin{eqnarray}
S &=& \sum _{j=1}^{N-1} a_j \int d^3 x \left\{ \frac{1}{2}(\partial_0 {A_z}_j - \phi'_j)^2 + \frac{1}{2 f_j} (F_{0a})_j^2
-\frac{f_j}{2} (\partial_a {A_z}_j - {A'_a}_j)^2\right. \nonumber \\
&&  -\frac{1}{4} (F_{ab})_j^2 +\frac{1}{z_j^2 f_j} | \partial_0 \psi_j -i {\phi}_j \psi_j |^2-\frac{f_j}{z_j^2} |\psi'_j
-i {A_z}_j \psi_j |^2 \nonumber \\
&& \left. -\frac{1}{z_j^2} |\partial_a \psi_j -i {A_a}_j \psi_j|^2 - \frac{1}{z_j^4} m^2 |\psi_j|^2 \right\} \, ,
\label{eq:latact}
\end{eqnarray}
where $f_j \equiv f(z_j)$.  We define derivatives in the discretized action such that
\begin{equation}
\phi_j' \equiv (\phi_{j+1} - \phi_j) /a_j \, ,
\end{equation}
and similarly for the other fields in the theory.  We now construct a
3D theory that reproduces Eq.~(\ref{eq:latact}), up to differences at the $j=1$ and $N$ boundaries. These
differences will be required for a proper holographic interpretation of the deconstructed theory.
\begin{figure}
    \centering
    \includegraphics[width=8cm,angle=0]{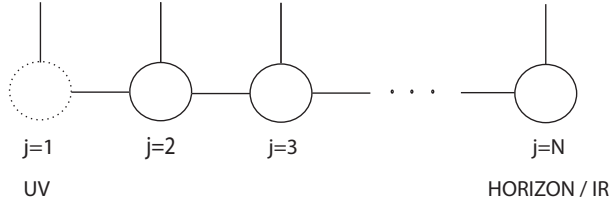}
    \caption{Moose diagram representation of the deconstructed theory. \label{fig:moose}}
\end{figure}

Consider a 3D theory with $N$ global U(1) symmetries of which $N-1$ are gauged.  We label
the $N-1$ gauge factors by the index $j=2 \dots N$.  We introduce a set of bi-fundamental
complex scalar fields $\Sigma_j$, $j=1\ldots N-1$, where $\Sigma_j$ transforms under the $j^{{\rm th}}$ and
$(j+1)^{{\rm th}}$ U(1) factors, with charges $-1$ and $+1$, respectively.  In addition, we introduce
 $N$ complex scalar fields $\psi_j$, $j=1 \ldots N$, where $\psi_j$ transforms under the
 $j^{{\rm th}}$ U(1) factor, with charge $+1$.  The particle content and charge assignments of this
 theory are conveniently summarized by the moose diagram shown in  Fig.~\ref{fig:moose}. The solid
 circles represent U(1) gauge groups, while each line connecting to a given circle represents a complex
 scalar field that transforms under that group.  The global U(1) symmetry is represented by the dotted
 circle at the left end of the moose.

We assume that the bi-fundamental fields develop vacuum expectation values (vevs) $\langle \Sigma_j \rangle \equiv v_j$.
Although the first link field $\Sigma_1$ is charged under the global U(1) factor, the global symmetry nevertheless
remains unbroken if the $\psi_j$ have vanishing expectation values.   
This can be seen by noting that the effect of a global phase rotation on $\langle \Sigma_1 \rangle$
can be undone by a (constant and spatially uniform) 
U(1) gauge transformation at the $j=2$ site.  The effect of this phase rotation on
$\langle \Sigma_2 \rangle$ can then be undone by a U(1) gauge transformation at the $j=3$ site, 
and so on.  In this way, one sees that the global symmetry of the first site remains an invariance of 
the vacuum, even when the link field vevs assume a  nonvanishing profile.  We therefore identify the 
current associated with the global symmetry of the first site as the QED current $j^\mu$.  As has been 
discussed in the literature~\cite{HHH}, even though the U(1) symmetry associated with QED is not 
gauged, the model still describes aspects of superconductivity to the
extent that a treatment of electromagnetic fields as external backgrounds is
appropriate.  In the deconstructed model we could simply add a kinetic term for
the associated U(1) gauge field to describe a fully dynamical 
electromagnetism, but in order to match the AdS/CFT description we will not do that here.

In order to compute expectation values of $j^\mu$, we will turn on a non-propagating 
gauge field at the first site, $A^\mu_1$, that couples to this current.   We will also assume 
that the gauge field at the last site $A^\mu_N$ is non-propagating and assumes a specified
background configuration; this will allow us to impose incoming-wave boundary conditions near the 
horizon.   The boundary fields $\psi_1$ and $\psi_N$ are also assumed to be 
non-propagating below, but only to simplify the discussion.   The case in which all the $\psi_j$ 
are dynamical is discussed in an appendix.

The Lagrangian for the 3D theory in Fig.~\ref{fig:moose} is of the form
\begin{equation}
{\cal L}=  \sum_{j=2}^{N-1} \left[- \frac{1}{4}(F_{\mu\nu})_j (F^{\mu\nu})_j
+Z_j | D_\mu \psi_j |^2 \right]+\sum_{j=1}^{N-1} \left[ |D_\mu \Sigma_j |^2 - Z_j V_j \right]
\label{eq:deconlag}
\end{equation}
where $V_j$ determines the scalar potential at the $j^{{\rm th}}$ site.  The coefficients $Z_j$
and the 3D metric $g^{\mu\nu}_j$ vary from site to site.  Although we use the term ``metric" to refer
to $g^{\mu\nu}_j$,  the $g^{\mu\nu}_j$ simply encode the Lorentz-violating couplings of the 
theory.  The covariant derivative is given by
\begin{equation}
D^\mu \Sigma_j = \partial^\mu \Sigma_j + i A^\mu_j \Sigma_j - i A^\mu_{j+1} \Sigma_j  \, ,
\end{equation}
and
\begin{equation}
D^\mu \psi_j = \partial^\mu \psi_j - i A^\mu_j \psi_j \, .
\end{equation}
Notice the absence of $\psi_j$ and $A_j$ kinetic terms for $j=1$ and $j=N$ in Eq.~(\ref{eq:deconlag}).  We identify
the non-propagating fields that appear at the ends of the moose with fixed backgrounds in the ultraviolet (UV)
and the infrared (IR):
\begin{equation}
\begin{array} {cccccc}  A^\mu_1&\equiv&A^\mu_{UV}(x^\nu)  \, , &  \psi_1 &\equiv& \psi_{UV}(x^\nu) \, , \\
A^\mu_N  &\equiv& A^\mu_{IR} (x^\nu) \, , & \psi_N &\equiv& \psi_{IR}(x^\nu) \, .
\end{array}
\end{equation}

We can now compare Eq.~(\ref{eq:deconlag}) to the Lagrangian obtained by latticizing the $z$ coordinate of the
continuum theory, Eq.~(\ref{eq:latact}).  Up to differences that vanish in the $a\rightarrow 0$ limit, one finds after some
algebra that the latticized theory is recovered if one chooses
\begin{equation}
g^{00}_j = \frac{1}{f_j}\,,  \,\,\,\,\,\,\,\,\,\, g^{ab}_j = -\delta^{ab}\,, \,\,\,\,\,\,\,\,\,\,
Z_j = \frac{1}{z_j^2}   \,\,\,\,\,\,\,\,\,\, \mbox{ and }  \,\,\,\,\,\,\,\,\,\, v_j = \frac{1}{a_j} \sqrt{\frac{f_j}{2}} \, ,
\end{equation}
the scalar potential
\begin{equation}
V_j = \left\{ \begin{array}{lll}
2  | v_j \psi_{j+1} -\psi_j \Sigma_j|^2 & \mbox{ for } &  j=1 \\
2  | v_j \psi_{j+1} -\psi_j \Sigma_j|^2 + m^2 |\psi_j|^2/z_j^2 & \mbox{ for } & j=2 \ldots N-1
\end{array} \right.
\label{eq:thepot}
\end{equation}
and the identification
\begin{equation}
\Sigma_j = v_j \exp[i  a_j {A_z}_j] \approx v_j (1+i a_j {A_z}_j) \, .
\label{eq:sigmap}
\end{equation}
The potential is not the most general one consistent with the symmetries of the theory; as in any deconstructed
model, the form of the action is dictated by the requirement that it reproduce the latticized action obtained from
the continuum theory.  For example, Eq.~(\ref{eq:thepot}) is fine-tuned so that the $(\psi_i')^2$ term
in Eq.~(\ref{eq:latact}) is reproduced when  $\Sigma_j$ is set equal to its vev.  For definiteness, we fix the mass
parameter $m^2$ to its holographically motivated value $m^2=-2$.  One could allow $m$ to deviate from this
choice in more general theories, but we will not consider that possibility here.  
The temperature dependence of terms in the action is inferred from the form of $f_j$, which depends on
temperature via $z_H = 3/(4 \pi T)$.  One could also imagine more general moose models in which the
temperature dependence of the $f_j$ is determined directly from the microscopic properties of the system. 
Here we will strictly consider the $f_j$ that follow from discretizing the continuum holographic theory.

In what follows, we work in unitary gauge, corresponding to the gauge $A_z=0$ in the four-dimensional theory.  
As in Eq.~(\ref{eq:sigmap}), we ignore the physical fluctuations of the link fields about their vevs. Then, the
Lagrangian of the moose model may be written
\begin{eqnarray}
{\cal L} &=& \sum _{j=1}^{N-1} a_j   \left[ \frac{1}{2}(\phi'_j)^2 -\frac{f_j}{2} ({A'_a}_j)^2 -\frac{f_j}{z_j^2} |\psi'_j |^2 \right]
+\sum _{j=2}^{N-1} a_j \left[ \frac{1}{2 f_j} (F_{0a})_j^2  -\frac{1}{4} (F_{ab})_j^2 \right] \nonumber \\
&+& \sum _{j=2}^{N-1} a_j  \left[ \frac{1}{z_j^2 f_j} | \partial_0 \psi_j -i {\phi}_j \psi_j |^2
-\frac{1}{z_j^2} |\partial_a \psi_j -i {A_a}_j \psi_j|^2 - \frac{1}{z_j^4} m^2 |\psi_j|^2 \right] \, .
\label{eq:mooseact}
\end{eqnarray}
One may now derive the equations of motion for the dynamical fields $(j=2 \ldots N-1)$, assuming the same ansatz
applied in the continuum theory.  For $\phi_j$ that are time-independent and spatially constant, the equations of motion
for the dynamical fields are given by
\begin{equation}
\phi''_j - \frac{2}{z_j^2 f_j} \phi_j \, \psi_j^2 =0 \, ,
\label{eq:phieq}
\end{equation}
where  $\phi''_j \equiv (\phi_{j+1}- 2\phi_j + \phi_{j-1})/a^2$, for $j=2 \ldots N-2$ and
$\phi''_{N-1} \equiv [(\phi_N-\phi_{N-1})/a_H - (\phi_{N-1}-\phi_{N-2})/a]/a_H$.
For the same ansatz, the equations of motion for the
$\psi_j$ are given by
\begin{equation}
\psi''_j + \frac{1}{a_j} \left(1 - \frac{z_j^2}{z_{j-1}^2} \frac{f_{j-1}}{f_j} \right) \psi'_{j-1} +\frac{1}{f_j^2} \phi_j^2 \psi_j \,
-\frac{m^2}{z_j^2 f_j} \,\psi_j =0 \, ,
\label{eq:psieq}
\end{equation}
where $\psi''_j$ is defined analogously to $\phi''_j$ and $\psi_j' \equiv (\psi_{j+1} - \psi_j) /a_j$.    When 
$\phi_j$ is nonvanishing, corresponding to a nonvanishing chemical potential and charge density, 
the coupling between $\phi_j$ and $\psi_j$ in Eq.~\eqref{eq:psieq} acts as a negative squared mass 
term for each Cooper pair field $\psi_j$, creating an instability that is enhanced for $z_j$ near
$z_H$ by the temperature-dependent factor of $1/f_j^2$.

Finally, we will require the equation of motion for the spatial components of the gauge field, assuming spatially 
constant fields and the time dependence
\begin{equation}
{A_a}_j(x^\mu) \equiv {A_a}_j e^{-i \omega t} \, .
\end{equation}
Again for $j=2 \ldots N-1$, one finds
\begin{equation}
{A''_a}_j + \frac{1}{a_j}\left(1 - \frac{f_{j-1}}{f_j} \right) {A'_a}_{j-1} + \left( \frac{\omega^2}{f_j^2} - 2 \frac{\psi_j^2}{z_j^2 f_j}\right) {A_a}_j =0 \, ,
\end{equation}
where $A''_j$ and $A'_j$ are defined analogously to $\psi''_j$ and $\psi'_j$.  The exciton fields $(A_a)_j$ are
excited collectively by the external electromagnetic field $(A_a)_1$.

The horizon boundary condition on Eqs.~(\ref{eq:phieq}) and (\ref{eq:psieq}) that correspond to those 
of the continuum theory are
\begin{equation}
\phi_N =0 \,  \,\,\,\,\, \mbox{  and  } \,\,\,\,\,  \psi'_{N-1} = \frac{2}{3 z_N} \psi_N \, .
\label{eq:phipsibc}
\end{equation}
If the $\psi_N$ field were dynamical at the last site, it turns out that the same boundary 
condition on $\psi$ would follow from the $\psi_N$ equation of motion in the continuum limit, as 
we discuss in the Appendix.  The boundary condition on the spatial components of $A_N$ can be 
chosen to reproduce the incoming-wave boundary conditions in the continuum limit, as we discuss 
in more detail below.

In the continuum holographic theory, physical quantities of interest are related to the values and 
derivatives of the fields at the UV boundary.  In the deconstructed theory, we find that similar 
relations apply to the expectation value of the charge density $\rho$ and the current density 
$j^a$, which are identified via their coupling to the background field $A_{UV}^\mu$:
\begin{equation}
j^\mu = i \frac{\delta}{\delta {A_{UV}}_\mu} \ln \left\{\int \prod_{j=2}^{N-1} {\cal D} A_j {\cal D} \psi_j\,  e^{i S} \right\} \, .
\label{eq:funcint}
\end{equation}
This reduces at tree-level to
\begin{equation}
j^\mu = -\frac{\delta}{\delta {A_{UV}}_\mu}   S_{cl}[A_{UV}]\, ,
\label{eq:computej}
\end{equation}
where the dependence on $A_{UV}$ arises since the fields are evaluated on their classical equations of motion.
In the linearized theory, one finds algebraically that $S_{cl}$ evaluates to a sum of surface terms 
\begin{equation}
S_{cl} = \int d^3x  \, [{\cal L}_{UV} + {\cal L}_{IR}] \, ,
\end{equation}
where the terms involving the exciton fields are given by
\begin{equation}
{\cal L}_{UV} = -\frac{1}{2a} \, \phi_1 (\phi_2 - \phi_1) + \frac{1}{2 a} f_1\, A_{a1} (A_{a2}-A_{a1}) \label{eq:uvsurface}
\end{equation}
and
\begin{equation}
{\cal L}_{IR} = -\frac{1}{2 a_H} f_{N-1}\, A_{aN} (A_{aN}-A_{a \,N-1}) \, ,
\label{eq:irsurface}
\end{equation}
using the boundary condition $\phi_N = 0$.  Before evaluating Eq.~(\ref{eq:computej}), we 
first arrange that the IR surface term vanishes. This is consistent with the approach in the 
continuum holographic theory, where the IR surface term is also present, but is simply 
discarded~\cite{Son-Starinets}.  To eliminate Eq.~(\ref{eq:irsurface}), we can add an {\em ad hoc} 
term to the Lagrangian such that the desired IR boundary condition is consistent with 
vanishing IR surface term.  Near the horizon, the ingoing-wave solution satisfies
Eq.~(\ref{eq:Afirstord}), which can be imposed as a boundary condition in
the deconstructed model in a number of ways.  We choose to approximate the
ingoing-wave boundary condition as
\begin{equation}
A_{a\,{N-n-1}}' = \frac{i \omega}{f_{N-n-1}} A_{a\,N-n} \, 
\label{eq:axdiscretebc}
\end{equation}
for some $n$, with $n\ll N$ in the continuum limit $N\rightarrow\infty$. 
If $n$ is taken too small, the factor of $1/f$ in Eq.~(\ref{eq:axdiscretebc}) is large and 
magnifies the discrepancy between the discretized and continuum boundary conditions.
Integrating from $j=n$ towards the horizon and solving the $\omega^2$-dependent equations of motion, 
one obtains a relation between 
$A_{a\,N}$ and $A_{a\,N-1}$ that generally leads to a nonvanishing surface 
term.  However, if we add a new term to the Lagrangian of the form 
\begin{equation}
{\cal L}_{H}=\xi\left[c(\omega^2)A_{a\,N}-A_{a\,N-1}\right]^2, \label{eq:newbndry}
\end{equation}
then the new IR surface term, evaluated on the solution to the equations of motion,
becomes a function of $c(\omega^2)$.  This function can be chosen so that the
surface term vanishes.  Eq.~(\ref{eq:newbndry}) can be expressed as a function of the 
gauge-invariant field strength tensors $F_{xz\,N-1}$ and $F_{0x\,N-1}$, and the function $c(\omega^2)$
may be interpreted as a function of $-\partial_t^2$, which ultimately acts on the non-dynamical
field $A_{aN}$.  The function $c(\omega^2)$ may be replaced by a polynomial approximation, valid at 
least over some finite range in $\omega$.  In this way, our discrete approximation to the ingoing-wave 
boundary condition, Eq.~(\ref{eq:axdiscretebc}), can be imposed $n$ sites away from the IR boundary, while
the IR surface term is arranged to vanish.  

The expectation value of the current thus depends only on the UV surface terms, as in 
the continuum holographic theory.  Each $j=2$ field in Eq.~(\ref{eq:uvsurface}) can be related 
to the corresponding boundary field by the discrete version of a bulk-to-boundary propagator.  
For example, we could write $\phi_j = B_{j1} \phi_1$, for some bulk-to-boundary propagator 
$B_{j1}$.  Then the charge density is found by computing
\begin{equation}
\rho = \frac{\delta}{\delta \phi_1} \left[-\frac{1}{2a} \phi_1^2 \,(B_{21} -1) \right] = -\frac{1}{a} \phi_1 \, (B_{21}-1) \,.
\end{equation}
The quantity $\phi_1 (B_{21} -1)/a$ is nothing more than what we would find by numerically solving the $\phi$
equation of motion and evaluating $\phi_1'$ as we defined it previously.  Hence, we conclude
\begin{equation}
\rho = - \phi_1 ' \, ,
\end{equation}
in agreement with the holographic prescription for computing the charge density in the continuum theory. By
the same reasoning, the current density following from Eq.~(\ref{eq:uvsurface}) is given by
\begin{equation}
j^a = -f_1 {A'_a}_1 \, ,
\label{eq:current}
\end{equation}
which also agrees with the usual holographic prescription in the continuum, where $f_1 \rightarrow 1$.
Since $j^0$ is proportional to the charge density operator, we identify $\phi_1$ with the chemical
potential
\begin{equation}
\mu = \phi_1 \, .
\end{equation}

The transition to the superconducting phase occurs when the $\psi_j$ develop a 
nonvanishing profile.  In the continuum limit, $\psi(z)$ takes the form $\psi(z)= \psi^{(1)}z\,
+\,\psi^{(2)}z^{2}$ near the UV boundary, so that the existence of the condensate is determined by 
nonvanishing first or second derivatives:
\begin{equation}
\psi^{(1)} \equiv \psi'(z=0) \neq 0 \,\,\,\,\, \mbox{ or } \,\,\,\,\, \psi^{(2)} \equiv \frac{1}{2} \psi''(z=0) \neq 0 \,,
\end{equation}
In the discretized theory, we use the analogous expression $\psi_{j}=\psi^{(1)} z_{j}\,+\,\psi^{(2)} z^{2}_{j}\,$ 
evaluated at the first two lattice sites ($z_1 = \epsilon$ and $z_2=\epsilon+a$) to define the 
operators \(\psi^{(1)}\) and \(\psi^{(2)}\),
\begin{equation}
\psi^{(1)} = \frac{\psi_{2}\, \epsilon^{2} - \psi_{1} (\epsilon + a)^{2}}
  {\epsilon^{2} (\epsilon + a) - \epsilon (\epsilon + a)^{2}}
  \quad {\rm and} \quad
\psi^{(2)} = \frac{\psi_{2}\, \epsilon - \psi_{1}(\epsilon + a)}
  {\epsilon (\epsilon + a)^{2} - \epsilon^{2}(\epsilon + a)} \, .
\end{equation}
These definitions have the appropriate continuum limit. As in Sec.~\ref{sec:cmodel}, we restrict 
ourselves to the case where $\psi^{(1)}=0$ and we define the order parameter to be 
$\langle O_{2}\rangle \equiv \sqrt{2} \psi^{(2)}$, following the conventions of Ref.~\cite{HHH}.
Although $\psi^{(1)}$ and $\psi^{(2)}$ do not have the same holographic interpretation in the
deconstructed theory away from the continuum limit, our definitions provide an accurate measure
of whether a nonvanishing profile of the $\psi_j$ has been dynamically generated.

\section{Results} \label{sec:results}
\begin{figure} [ht]
  \centering
  \includegraphics[scale=.6]{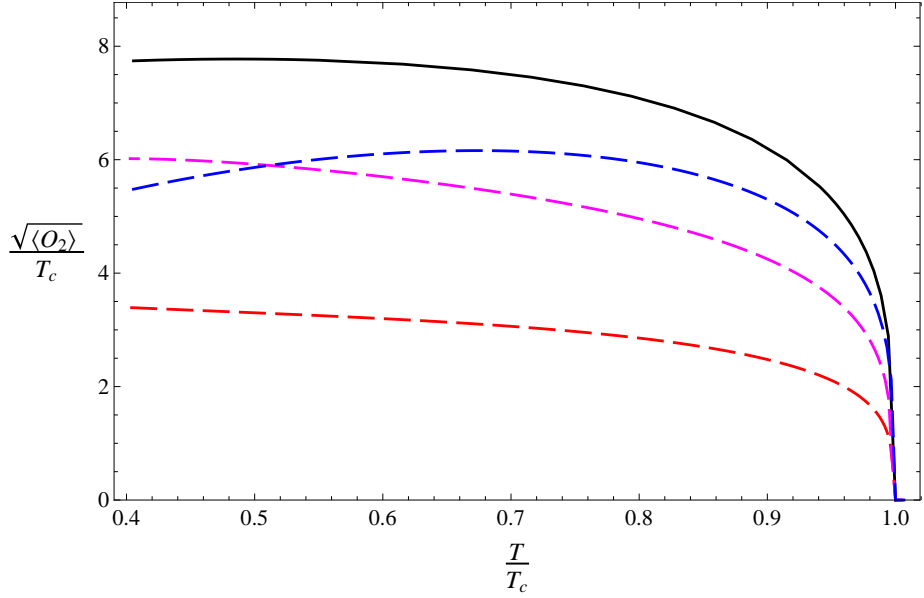}
  \caption{This figure displays the condensation phase transition at critical
    temperature $T = T_{\rm c}$. The top curve is the $N=1000$ curve
    and below that the dashed curves from top to bottom around $T_c$ 
    are $N=5$, $N=10$, and $N=100$, respectively.  The 1000-site curve 
    includes a fixed UV cutoff $\epsilon$, which allows for a smooth continuum limit.}
  \label{fig:condensates}
\end{figure}

In this section, we numerically solve the discretized theory of
Sec.~\ref{sec:moose} for \(N \in \left\{5, 10, 100, 1000\right\}\).
In the case of \(N=1000\), for the purpose of connecting with the continuum
theory, we fix the UV cutoff \(z_1 = \epsilon = 10^{-1}\),
while for other \(N\) we set \(\epsilon = a\), the
bulk lattice spacing.  We fix the lattice spacing at the horizon $a_H=10^{-5}$.
In all cases, we impose the following boundary conditions:
\begin{equation}
\phi_{1}' = -\rho = -1\, ,\quad \psi^{(1)} = 0\, ,\quad
  \phi_{N} = 0\, ,\quad {\rm and} \,\,
  \psi_{N-1}' = \frac{2}{3 z_{N}} \psi_{N}\, ,
\end{equation}
where we have used the scaling symmetry of the theory to fix $T_c$ by setting $\rho=1$.
Searching for nonvanishing \(\psi_{j}\) solutions as we vary the temperature,
we find there is a critical temperature for each $N$
at which the operator \(\langle O_{2}\rangle\) condenses, as shown in
Fig.~\ref{fig:condensates}. We note that there are numerous non-zero
solutions for \(\psi_{j}\) but, as in the continuum model~\cite{HHH}, we retain
only the monotonic solutions.  
For \(N = 5\), \(N = 10\), \(N = 100\), and
\(N = 1000\) we find the critical temperatures \(T_{\rm c} = 0.076\,\rho^{1/2}\),
\(T_{\rm c} = 0.088 \, \rho^{1/2}\), \(T_{\rm c} = 0.111 \, \rho^{1/2}\), and
\(T_{\rm c} = 0.118 \, \rho^{1/2}\) respectively.
Here we reintroduced the factor of \(\rho^{1/2}\) as indicated by the scaling
relations. It is interesting to see that for all \(N\) we retain the square root
behavior of the phase transition. We find that
the curves in Fig.~\ref{fig:condensates} near $T_c$ are well fit by the
form \(\langle O_{2} \rangle = C_{N} T^{2}_{\rm c}
(1 - T/T_{\rm c})^{1/2}\).  The values we obtain for the coefficients are
\(C_{1000} = 127,\,C_{100} = 20,\,C_{10} = 58,\) and \(C_{5} = 93\),
as compared to the continuum value $C=144$~\cite{HHH}.

\begin{figure}[ht]
\subfigure{\includegraphics[width = 0.495\textwidth]{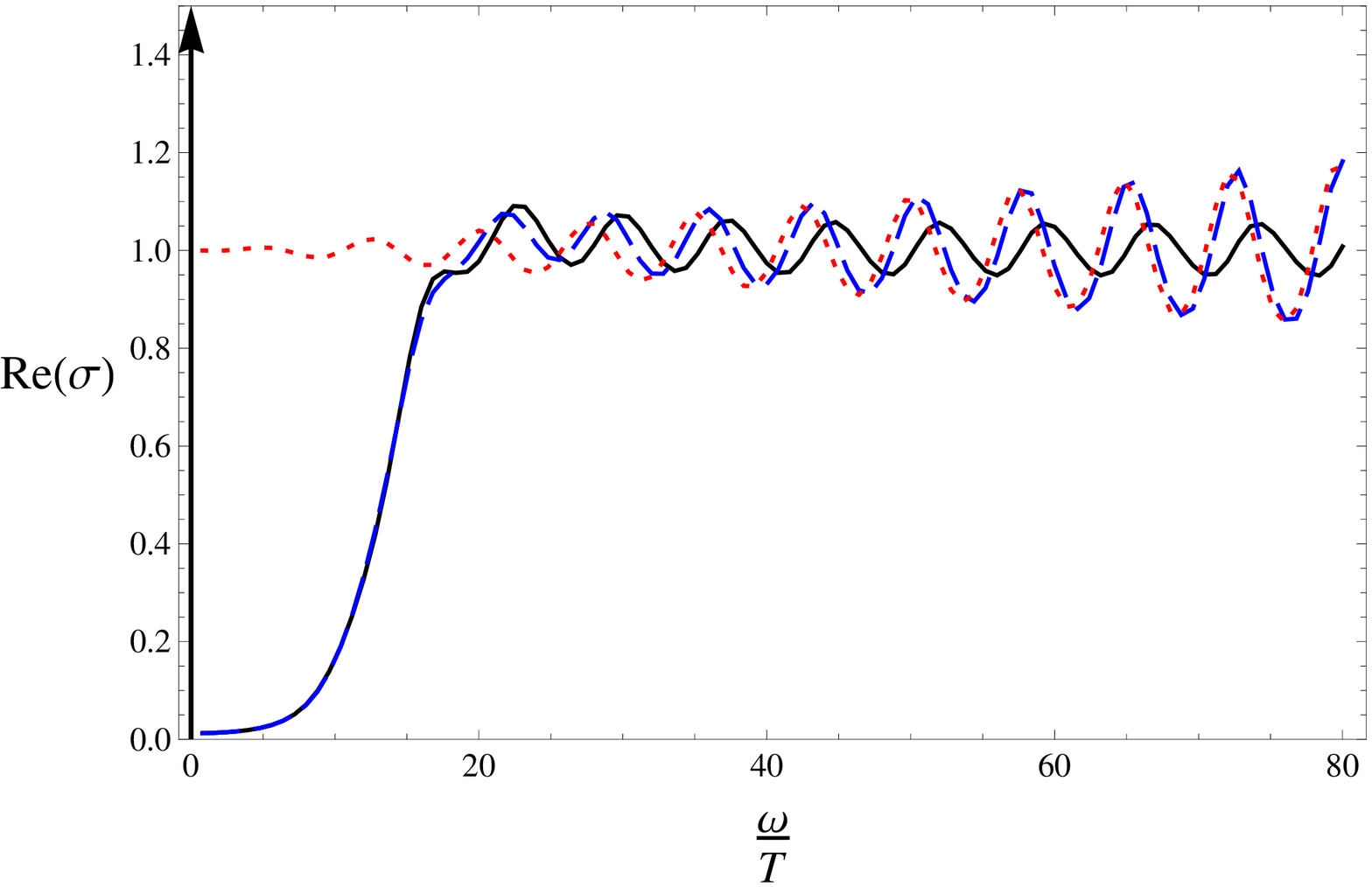}}
\subfigure{\includegraphics[width = 0.495\textwidth]{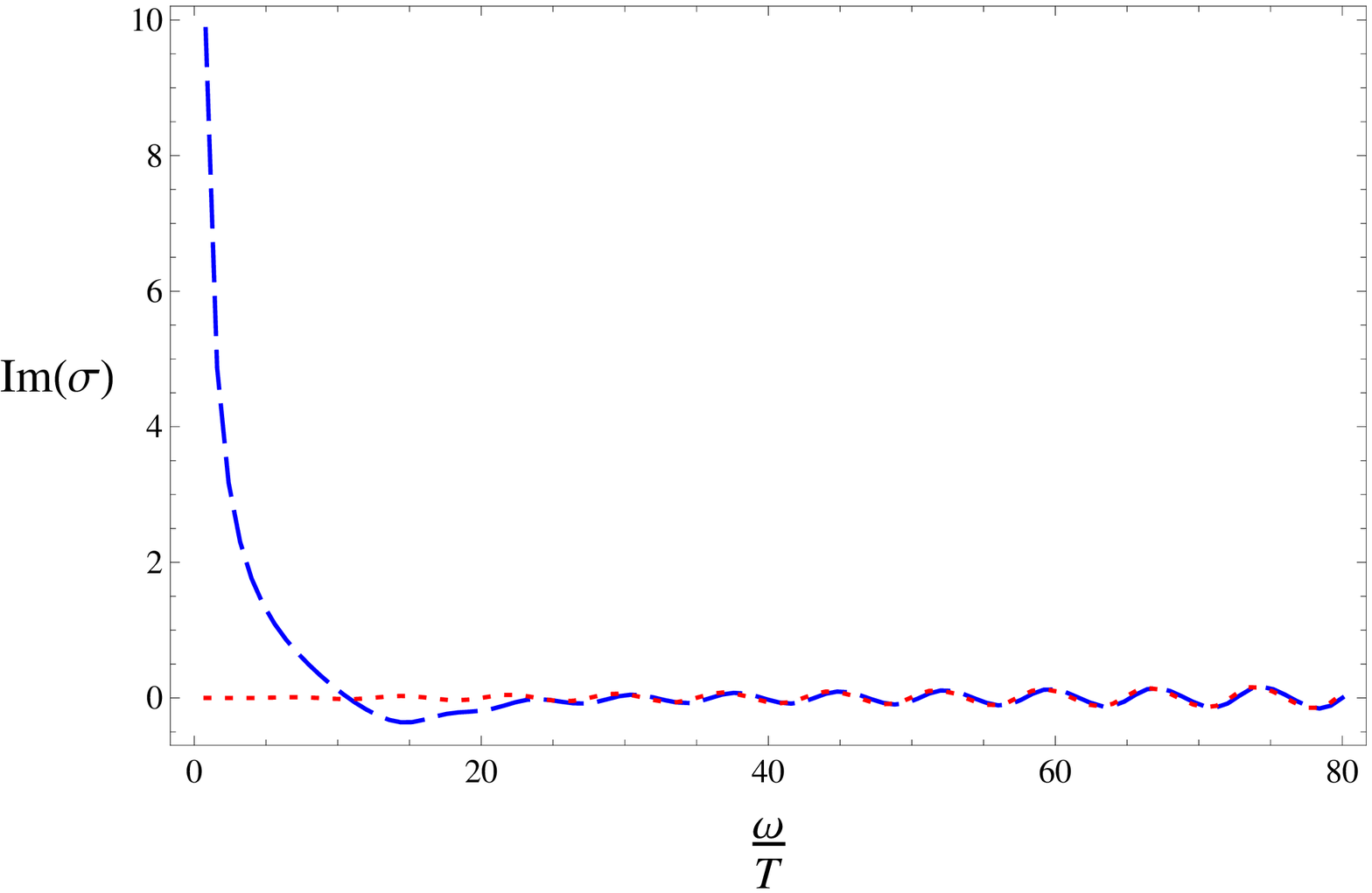}}
\subfigure{\includegraphics[width = 0.495\textwidth]{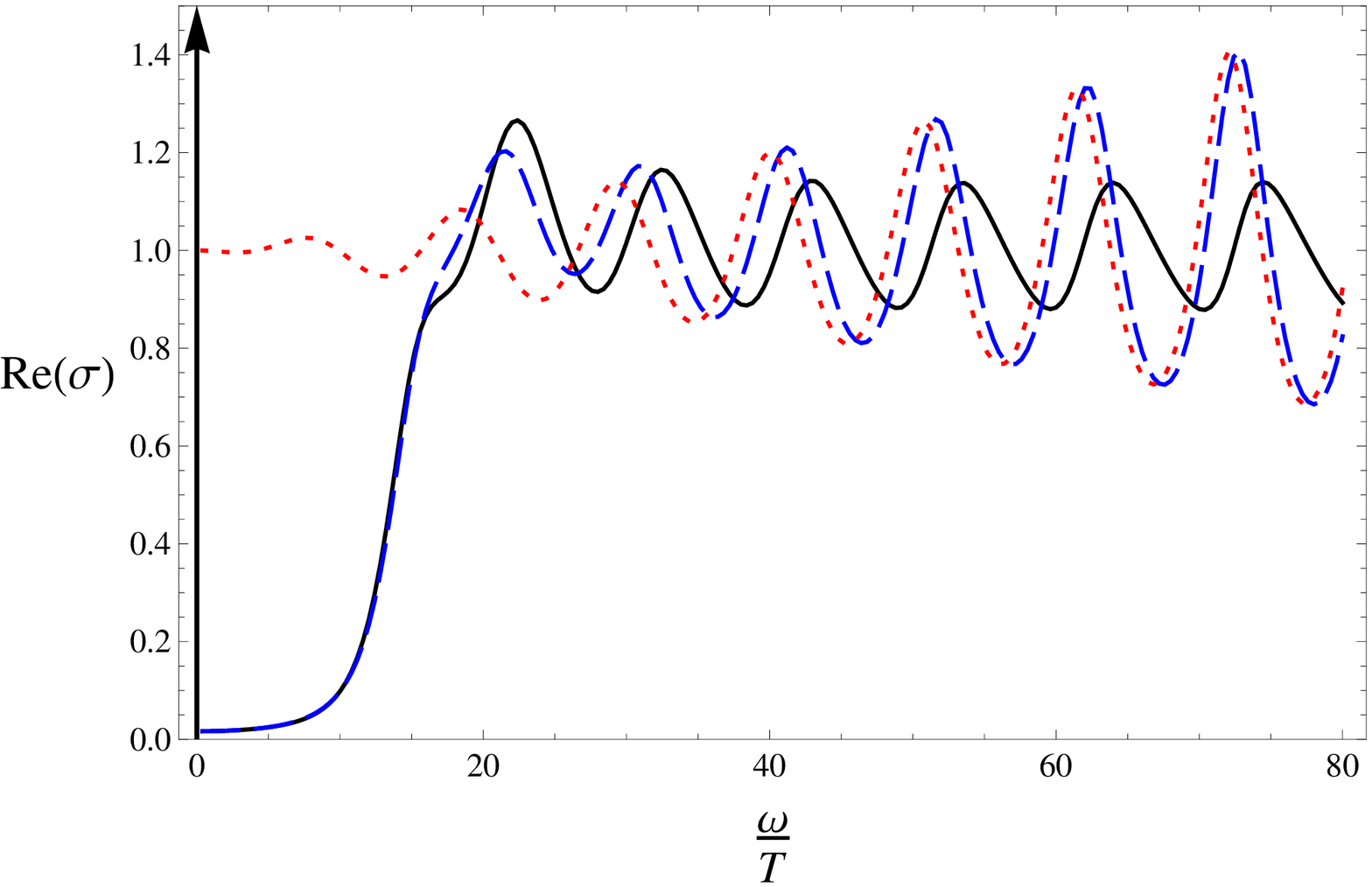}}
\subfigure{\includegraphics[width = 0.495\textwidth]{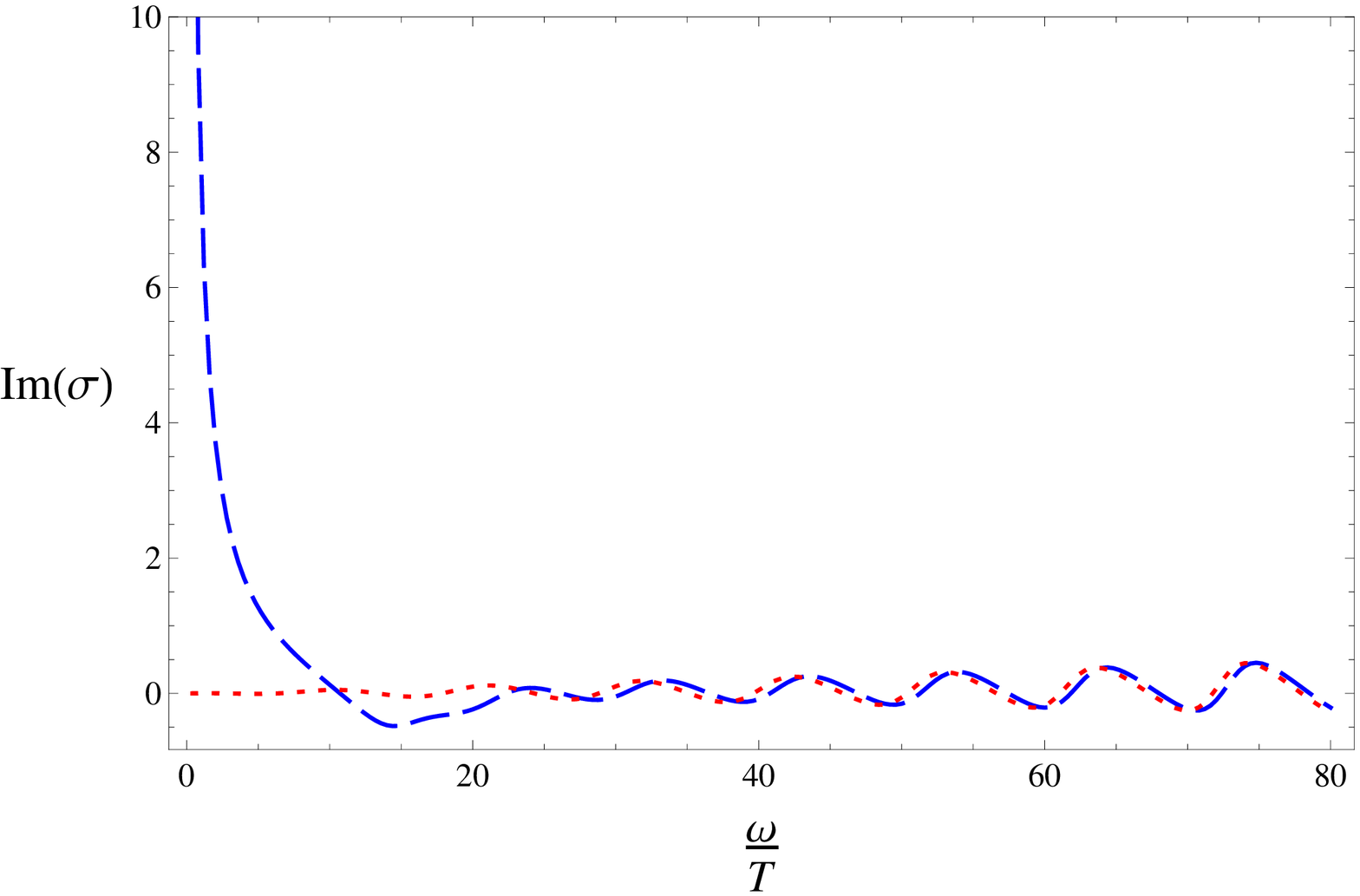}}
\caption{
  Complex conductivities.  The $(N,T/T_{\rm c})$ values for the top and bottom rows are given by 
  $(1000, 0.55)$ and $(100, 0.53)$, respectively.   The dotted (red) curve is the normal (n) 
  phase while the dashed (blue) curve is the superconducting (sc) phase.  The vertical arrow represents a
    delta function. The solid (black) curve is the ratio: ${\rm Re}(\sigma_{\rm sc})/{\rm Re}(\sigma_{\rm n})$. }
\label{fig:conductivities1}
\end{figure}

\begin{figure}
\subfigure{\includegraphics[width = 0.495\textwidth]{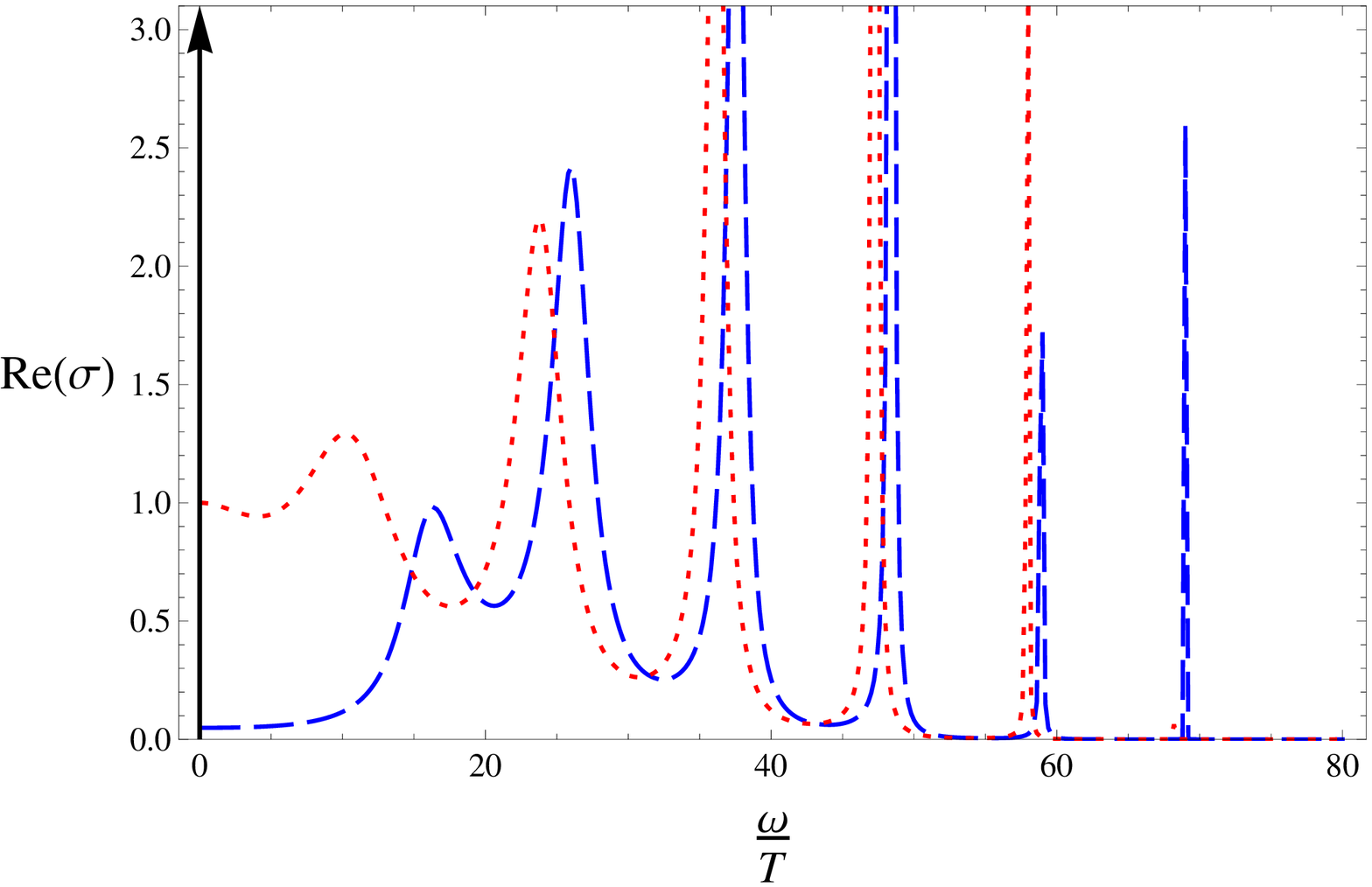}}
\subfigure{\includegraphics[width = 0.495\textwidth]{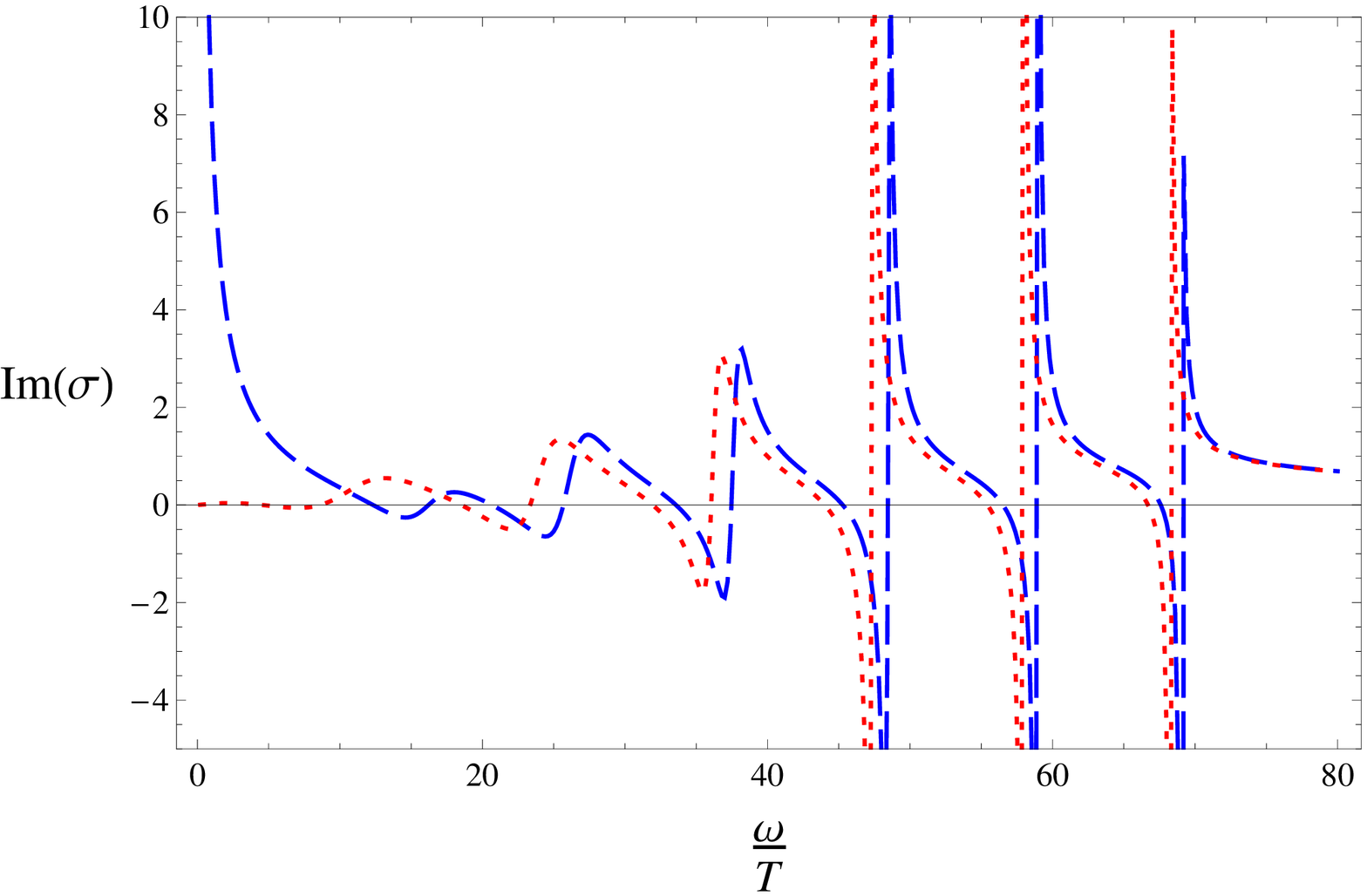}}
\subfigure{\includegraphics[width = 0.495\textwidth]{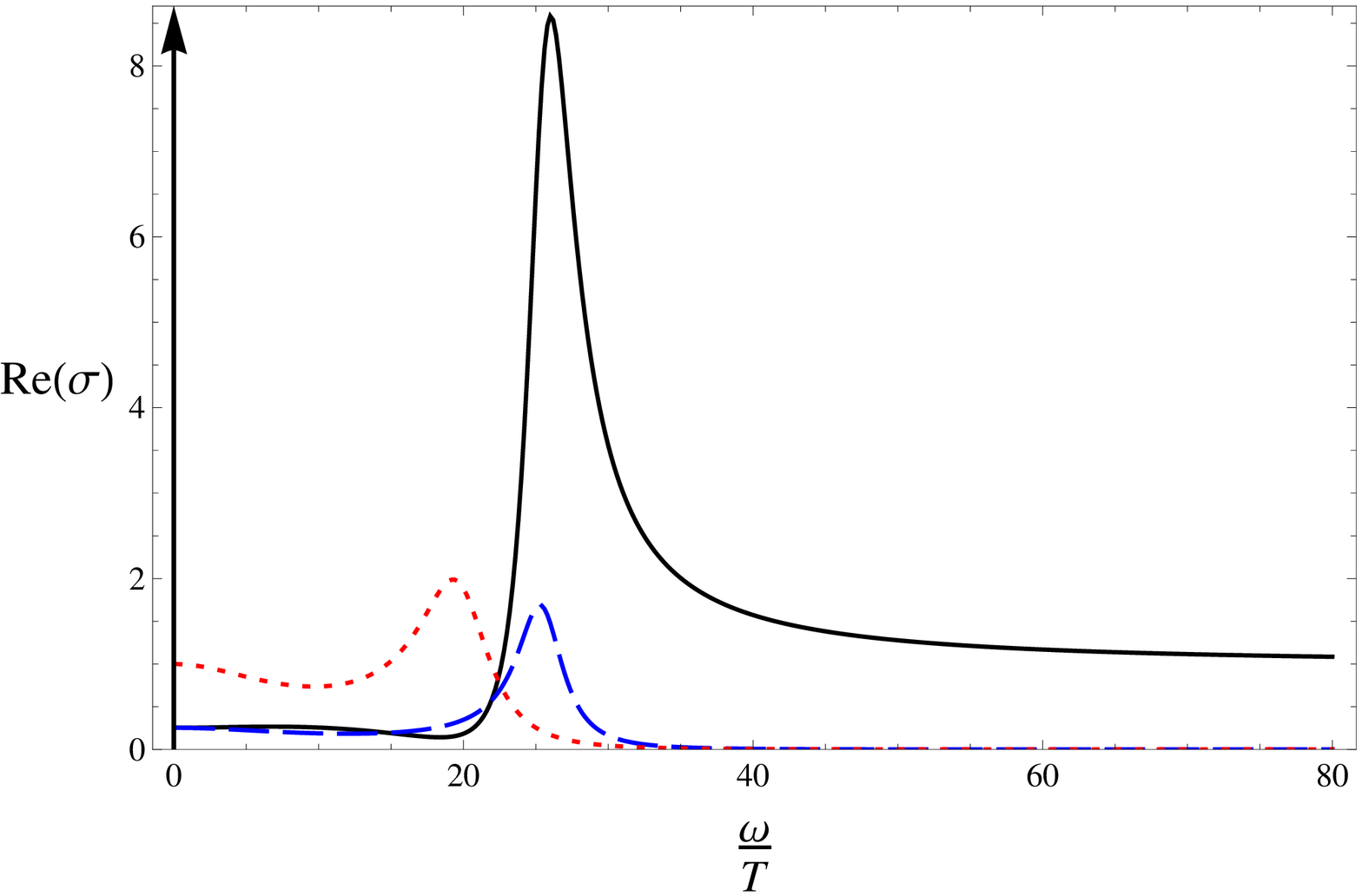}}
\subfigure{\includegraphics[width = 0.495\textwidth]{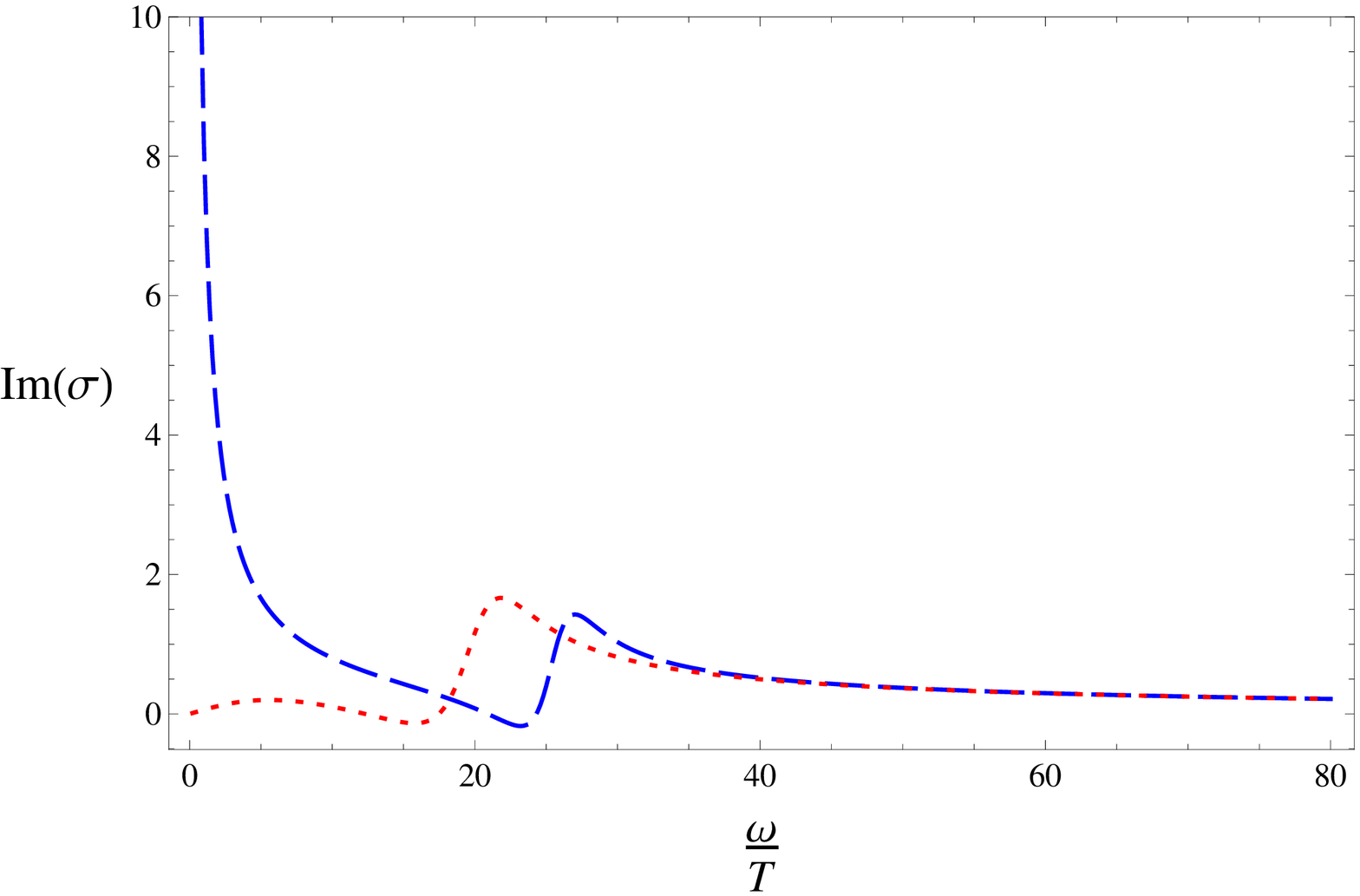}}
  \caption{
  Complex conductivities.  The $(N,T/T_{\rm c})$ values for the top and bottom rows are given by 
  $(10,0.54)$ and $(5, 0.50)$, respectively.  The dotted (red) curve is the normal (n) phase 
  while the dashed (blue) curve is the superconducting (sc) phase.  The vertical arrow represents a
  delta function. The solid (black) curve (omitted for $N=10$, for clarity) is the 
  ratio: ${\rm Re}(\sigma_{\rm sc})/{\rm Re}(\sigma_{\rm n})$. }
  \label{fig:conductivities2}
\end{figure}

We would also like to study the behavior of the complex conductivity away from the continuum
limit.
Using the expression for the current Eq.~\eqref{eq:current}, we
find the following expression for the conductivity:
\begin{equation}
\sigma = \frac{j^{x}}{E_{x}} = \frac{j^{x}}{\dot{A_{x}}}
  = -\frac{i f_1(A_{x2} - A_{{\rm UV}x})/a}{\omega A_{{\rm UV}x}}.
\end{equation}
As discussed in Sec.~\ref{sec:moose}, we impose the
ingoing-wave boundary condition, Eq.~(\ref{eq:axdiscretebc}), $n$ sites away from the horizon.
In particular, we set
\begin{equation}
A_{x\, N - n} = 1 \quad {\rm and} \quad A_{x\, N - n - 1} = 1 -
  \frac{i \omega a}{f_{N - n - 1}} \, ,
\end{equation}
where we have fixed the arbitrary normalization \(A_{x\, N-n} = 1\).
Our choices for the shift \(n\) for \(N = 1000, 100, 10\) and \(5\) are
\(n = 20,10,2\) and \(2\), respectively.  With these boundary
conditions for \(A_{xj}\) we compute the real and imaginary parts of \(\sigma\), as shown in
Figs.~\ref{fig:conductivities1} and \ref{fig:conductivities2}.  It is important to note that the graphs of
\({\rm Im}[\sigma(\omega)]\) in the superconducting phase exhibit a simple pole at
\(\omega = 0\). This implies the presence of a delta
function contribution to the real part, \({\rm Re}[\sigma(\omega)] \propto \delta(\omega)\),
as discussed in Sec.~\ref{sec:cmodel}.
For \(N = 10\) there are a series of peaks in \({\rm Re}[\sigma(\omega)]\) with a corresponding 
pole-like structure in the imaginary part as follows from the Kramers-Kronig relation 
Eq.~(\ref{eq:Kramers-Kronig}).  Finally, the oscillatory behavior of ${\rm Re}[\sigma(\omega)]$ at 
large $\omega$ displayed in the $N=100$ and $N=1000$ plots is consistent with our expectation 
from Eq.~(\ref{eq:wiggles}) given that our boundary condition is only approximately ingoing wave.


\section{Conclusions} \label{sec:conclusions}

Dimensional deconstruction is a powerful technique for finding lower-dimensional theories that share
some of the interesting phenomenological features of their higher-dimensional progenitors.  In this paper,
we have considered the dimensional deconstruction of four-dimensional holographic theories of superconductivity.
We have shown how the AdS/CFT prescriptions for computing physical quantities of interest (for example, charge
densities and currents) emerge in the lower-dimensional, discretized theory.   We have also demonstrated that
deconstructed theories with a relatively small number of lattice sites (here taken $\geq 5$) do indeed retain enough
of the physics of the continuum limit to provide possible models of superconductivity on their own.

We view the results of this work as preliminary, in the sense that we have only considered the deconstruction of one
of the simplest, s-wave holographic superconductors that were proposed in the first papers on the application
of the AdS/CFT correspondence to this problem.  Much of what we have found in the present work leads to questions 
and suggestions for future study:

\begin{itemize}
\item Dimensional deconstruction also motivates new models of $p$- and $d$-wave superconductors.
It would be interesting to study the phenomenology of those models.

\item In the present approach, the photon field is not included as a dynamical degree of freedom in the
deconstructed theory.  However, the theory can be easily modified by gauging the electromagnetic U(1) symmetry 
and adding an associated gauge kinetic term.

\item The deconstructed model suggests the possibility of hidden local symmetries in the interactions between
Cooper pairs and excitons.  Perhaps absorption studies would be able to test the existence of these states
and interactions.

\item If one deviates from a strict matching to the continuum holographic theory, then one has greater freedom in
constructing a model that is based on a very small number of replicated gauge groups (for example, two or three).
Would a  two- or three-site model of superconductivity be phenomenologically viable?

\item The deconstructed model introduces temperature-dependent couplings inferred by a holographic model.  It
would be important to deduce these couplings instead from a microscopic description.

\item In the present work, the fluctuations of the link fields were ignored, consistent with the tacit assumption that
these fields are heavy.  However, these fields, which have no counterpart in the continuum holographic theory,
might also have a physical interpretation.
\end{itemize}

We look forward to considering these issues in more detail in future work.

\begin{acknowledgments}
We thank Henry Krakauer, Enrico Rossi and Christopher Triola for useful conversations.  
This work was supported by the NSF under Grant PHY-1068008.  In addition, C.D.C. thanks Joseph J. 
Plumeri II for his generous support.
\end{acknowledgments}

\appendix
\section{A purely dynamical $\psi$}
To simplify our discussion, it was assumed earlier that the $\psi_j$ fields at the ends of the 
moose, $\psi_1$ and $\psi_N$, were background fields.  Here we show that the same boundary conditions 
would be obtained in the continuum limit if we allow all the $\psi_j$ to be dynamical fields.   The Lagrangian 
that is relevant in this case is a minor modification of Eq.~(\ref{eq:mooseact});
the third sum in this expression, which runs from $j=2$ to $N-1$, is changed to one which 
runs from $j=1$ to $N$ (with $a_N=a_H$) for the first two
terms in square brackets and from $j=2$ to $N$ for the mass squared term. We also 
shift $z_N$ away from its previous value by $\epsilon_H$, which serves
 as a cutoff to regulate the divergence in $1/f_N$.
Eq.~(\ref{eq:psieq}) remains unaltered, but now there are two additional equations of 
motion, for $\psi_1$ and $\psi_N$:
\begin{equation}
\frac{1}{a} \psi'_1+ \frac{1}{f^2_1} \phi_1^2 \psi_1 =0 \, ,
\label{eq:uveom}
\end{equation}
and
\begin{equation}
-\frac{1}{a_H} \psi'_{N-1}+\frac{z_{N-1}^2 }{z_N^2 f_N f_{N-1}} \phi_N^2 \psi_N 
+ \frac{2 z_{N-1}^2}{f_{N-1} z_N^4} \psi_N =0 \,,
\label{eq:ireom}
\end{equation}
where $\psi'_{N-1} = (\psi_N - \psi_{N-1})/a_H$.  These equations can be viewed as 
dictating the boundary conditions for Eq.~(\ref{eq:mooseact}), which, in
our previous approach, were chosen freely to mimic the boundary conditions of 
the continuum theory. Eq.~(\ref{eq:uveom}) reduces in the continuum limit to the
boundary condition $\psi'_1=0$.  Since
$\phi_N \equiv 0$, the second term in Eq.~(\ref{eq:ireom}) vanishes ($1/f_N$ is 
finite for finite cutoff $\epsilon_H$).  Hence,
\begin{equation}
\psi'_{N-1} = \left[ \frac{2 a_H z_{N-1}^2}{f_{N-1} z_N^4} \right] \psi_N\, .
\label{eq:newpsihbc}
\end{equation}
The quantity in brackets can be expanded in powers of $a_H$ while maintaining $\epsilon_H\ll a_H$; one finds
\begin{equation}
\psi'_{N-1} = \left[\frac{2}{3 z_H} - \frac{2}{3 z_H^2} a_H + {\cal O} (a_H^2)\right] \psi_N \, .
\end{equation}
In the limit $a_H \rightarrow 0$ we recover the boundary condition  in Eq.~(\ref{eq:phipsibc}).
In summary, the effect of treating the $\psi_j$ as dynamical fields everywhere is to modify the form of the
boundary conditions away from those of the continuum limit:
\begin{equation}
\psi'_1= - \frac{a}{f^2_1} \phi_1^2 \psi_1 \, , \,\,\,\,\, \mbox{ and } \,\,\,\,\, \psi'_{N-1}= \left[ \frac{2}{3 z_H} 
-\frac{2}{3z_H^2}a_H\right] \psi_N \, .
\label{eq:altbc}
\end{equation}

\end{document}